\documentclass[10pt,a4paper]{article}
\usepackage[intlimits]{amsmath}
\usepackage[psamsfonts]{amssymb}
\usepackage{layout}
\usepackage{enumerate}
\usepackage{longtable}
\usepackage{array}

\allowdisplaybreaks
 \allowdisplaybreaks

\begin{document}

Quantum Computers and Computing, 2006, v. 6, n. 1, p. 125-136.

\begin{center}
{\large \textbf{Probabilistic Simulation of Quantum Computation}}\\[10mm]
\textit{T.F. Kamalov$^1$ and Yu.P. Rybakov$^2$}\\[0pt]
\textit{${}^1$Physics Department, Moscow State Opened University\\[0pt]
${}^2$Theoretical Physics Department, Peoples' Friendship University of
Russia}\\[0pt]
\textit{E-mail: ${}^1$ykamalov@rambler.ru, qubit@mail.ru\\[0pt]
${}^2$soliton4@mail.ru}\\[0pt]
\begin{abstract}
{Special stochastic representation of the wave function in Quantum
Mechanics (QM), based on soliton realization of extended
particles, is suggested with the aim to model quantum states via
classical computer. Entangled solitons construction being
introduced in the nonlinear spinor field model, the
Einstein---Podolsky---Rosen (EPR) spin correlation is calculated
and shown to coincide with the quantum mechanical one for the
spin--$1/2$ particles in the singlet state. The concept of
stochastic qubits is used for quantum computing modelling.}
\end{abstract}

\textit{Keywords:} stochastic simulation of qubits, entangled solitons,
random Hilbert space.\\[0pt]
\end{center}

{\Large PACS:} 03.65.Ud\newline

\section{Geometric Quantum Mechanics and its Stochastic Representation}

Recent years a very fascinating idea to put QM into geometric language
attracts the attention of many physicists~[1]. The starting point for such
an approach is the projective interpretation of the Hilbert space $\mathcal{H%
}$ as the space of rays. To illustrate the main idea it is convenient to
decompose the Hermitian inner product $\langle \cdot |\cdot \rangle $ in $%
\mathcal{H}$ into real and imaginary parts by putting for the two $L_{2}$%
--vectors $|\psi _{1}\rangle =u_{1}+\imath v_{1}$ and $|\psi _{2}\rangle
=u_{2}+\imath v_{2}$:
\begin{equation}
\langle \psi _{1}|\psi _{2}\rangle =G\,(\psi _{1},\psi _{2})-\imath \Omega
\,(\psi _{1},\psi _{2}),  \label{eq:1}
\end{equation}%
where $G$ is a Riemannian inner product on $\mathcal{H}$ and $\Omega $ is a
symplectic form, that is
\begin{equation*}
G\,(\psi _{1},\psi _{2})=(u_{1},u_{2})+(v_{1},v_{2});\quad \Omega \,(\psi
_{1},\psi _{2})=(v_{1},u_{2})-(u_{1},v_{2}),
\end{equation*}%
with $(\cdot ,\cdot )$ denoting standard $L_{2}$ inner product. The
symplectic form $\Omega $ revealed in (1) can acquire its dynamical content
if one uses the special stochastic representation of QM suggested in~[2, 3,
4].

As a first motivation for introducing stochastic representation of the wave
function let us consider the de Broglie plane wave
\begin{equation}  \label{eq:2}
\psi = A\,e^{-\imath kx} = A\,e^{-\imath \omega t+\imath(\mathbf{kr})}
\end{equation}
for a free particle with the energy $\omega $, momentum $\mathbf{k}$, and
mass $m$, when the relativistic relation
\begin{equation*}
k^2 = \omega^2-{\mathbf{k}}^2 = m^2
\end{equation*}
holds (in natural units $\hbar = c = 1$).

Suppose, following L.~de~Broglie~[5] and A.~Einstein~[6], that the structure
of the particle is described by a regular bounded function $u(t, \mathbf{r})$%
, called hereafter as soliton, which is supposed to satisfy some nonlinear
equation with the Klein---Gordon linear part. Let ${\ell }_0 = 1/m$ be the
characteristic size of the soliton solution $u(t, \mathbf{r})$ moving with
the velocity $\mathbf{v} = {\mathbf{k}}/{\omega}$.

Now it is worth-while to underline the remarkable fact behind this
research~[7], namely, the possibility to represent the de Broglie wave (\ref%
{eq:2}) as the sum of solitons located at nodes of a cubic lattice with the
spacing $a\gg {\ell }_{0}$:
\begin{equation}
A\,e^{-\imath kx}=\sum_{\mathbf{d}}u(t,\mathbf{r}+\mathbf{d}),  \label{eq:3}
\end{equation}%
where $\mathbf{d}$ marks the positions of lattice nodes. To show the
validity of (3) one can take into account the asymptotic behavior of the
soliton in its tail region:
\begin{equation*}
u(x)=\int \,d^{4}k\,e^{-\imath kx}g(k)\delta (k^{2}-m^{2})
\end{equation*}%
and then use the well-known formula
\begin{equation*}
\sum_{\mathbf{d}}e^{\imath (\mathbf{k}\,\mathbf{d})}={\left( \frac{2\pi }{a}%
\right) }^{3}\delta (\mathbf{k}),
\end{equation*}%
implying that
\begin{equation*}
A={\left( \frac{2\pi }{a}\right) }^{3}\frac{g(m)}{2m}.
\end{equation*}%
The formula (3) gives a simple illustration of the wave---particle dualism,
showing that the de Broglie wave characterizes the assemblage of
particles---solitons.

The shortest way to get the stochastic representation of quantum mechanics
is modify the formula (3). This can be easily performed if one admits that
the locations of solitons' centers are not regular nodes of the cubic
lattice but some randomly chosen points. To realize this prescription,
suppose that the real field~$\phi $ describes $n$ particles---solitons and
has the form
\begin{equation}
\phi (t,\mathbf{r})=\sum_{k=1}^{n}\phi ^{(k)}(t,\mathbf{r}),  \label{eq:4}
\end{equation}%
where
\begin{equation*}
\mbox{supp}\,\phi ^{(k)}\,\cap \,\,\mbox{supp}\,\phi ^{(k^{\prime
})}=0,\quad k\neq k^{\prime },
\end{equation*}%
and the same for the conjugate momenta
\begin{equation*}
\pi (t,\mathbf{r})=\partial \mathcal{L}/\partial \phi _{t}=\sum_{k=1}^{n}\pi
^{(k)}(t,\mathbf{r}),\quad \phi _{t}=\partial \phi /\partial t,
\end{equation*}%
where $\mathcal{L}$ stands for the Lagrangian density of the field $\phi $.
Let us define the auxiliary functions
\begin{equation}
\varphi ^{(k)}(t,\mathbf{r})=\frac{1}{\sqrt{2}}(\nu _{k}\phi ^{(k)}+\imath
\pi ^{(k)}/\nu _{k})  \label{eq:5}
\end{equation}%
with the constants $\nu _{k}$ satisfying the normalization condition
\begin{equation}
\hbar =\int \,d^{3}x\,|\varphi ^{(k)}|^{2}.  \label{eq:6}
\end{equation}%
Now we define the analog of the wave function in the configurational space $%
\mathbb{R}^{3n}\ni \mathbf{x}=\{\mathbf{r}_{1},\ldots ,\mathbf{r}_{n}\}$ as
\begin{equation}
\Psi _{N}(t,\mathbf{r}_{1},\ldots ,\mathbf{r}_{n})=(\hbar
^{n}N)^{-1/2}\sum_{j=1}^{N}\prod_{k=1}^{n}\varphi _{j}^{(k)}(t,\mathbf{r}%
_{k}),  \label{eq:7}
\end{equation}%
where $N\gg 1$ stands for the number of trials (observations) and $\varphi
_{j}^{(k)}$ is the one-particle field function for the $j$--th trial.

Now we intend to show that the quantity
\begin{equation*}
\rho _{N}=\frac{1}{(\triangle \vee )^{n}}\,\int\limits_{(\triangle \vee
)^{n}\subset \mathbb{R}^{3n}}d^{3n}x\,\left\vert \Psi _{N}\right\vert ^{2},
\end{equation*}%
where $\triangle \vee $ is the elementary volume which is supposed to be
much greater than the proper volume of the particle ${\ell _{0}}^{3}=\vee
_{0}\ll \triangle \vee $, plays the role of the coordinate probability
density. To this end let us calculate the following integral:
\begin{equation*}
\left( \triangle \vee \right) ^{n}\rho _{N}\equiv \int\limits_{\left(
\triangle \vee \right) ^{n}}{d^{3n}x\,\left\vert {\Psi _{N}}\right\vert ^{2}}%
=\left( {\hbar ^{n}N}\right) ^{-1}\left( {\sum\limits_{i=1}^{N}{a_{ii}}%
+\sum\limits_{i\neq j=1}^{N}{a_{ij}}}\right) ,
\end{equation*}%
where the denotation is used
\begin{equation*}
a_{ij}=\textstyle{\frac{1}{2}}\prod\limits_{k=1}^{n}{\int\limits_{\triangle
\vee }{d^{3}x\,\left( {\varphi _{i}^{\ast (k)}\varphi _{j}^{(k)}+\varphi
_{j}^{\ast (k)}\varphi _{i}^{(k)}}\right) }}.
\end{equation*}%
Taking into account (6) one gets
\begin{equation}
\left( \triangle \vee \right) ^{n}\rho _{N}=\left( {\hbar ^{n}N}\right)
^{-1}\left( {\hbar ^{n}\triangle N+S}\right) ,\quad S=\sum\limits_{i\neq j}{%
a_{ij}},  \label{eq:8}
\end{equation}%
with $\triangle N$ standing for the number of trials for which the centers
of particles---solitons were located in $\left( {\triangle \vee }\right) ^{n}
$. It is worth-while to remark that due to independence of trials and
arbitrariness of initial data and, in particular, of the phases of the
functions $\varphi _{i}^{(k)}$, one can consider the entities $a_{ij}$ for $%
i\neq j$ as independent random variables with zero mean values. This fact
permits to use the Chebyshev's inequality~[8] to estimate the probability of
the events for which $|S|$ surpasses $\hbar ^{n}\triangle N$:
\begin{equation}
P\left( {\left\vert S\right\vert >\hbar ^{n}\triangle N}\right) \leq \left( {%
\hbar ^{n}\triangle N}\right) ^{-2}\left\langle {S^{2}}\right\rangle .
\label{eq:9}
\end{equation}%
On the other hand, in view of trials' independence one gets
\begin{equation}
\left\langle {S^{2}}\right\rangle =\sum\limits_{i\neq j}{\left\langle {%
a_{ij}^{2}}\right\rangle }.  \label{eq:10}
\end{equation}%
Now one can take into account that the wave packets $\varphi _{i}^{(k)}$ are
effectively overlapped if their centers belong to the proper volume domain $%
\vee _{0}$. This property permits to deduce from (6) and (10) the estimate
\begin{equation}
\left\langle {S^{2}}\right\rangle \leq \alpha ^{n}\hbar ^{2n}\frac{\triangle
N}{(\triangle \vee )^{n}}{\vee _{0}}^{n}\triangle N,  \label{eq:11}
\end{equation}%
where $\alpha \sim 1$ is the \textquotedblleft packing" factor for the
nearest neighbors. Inserting (11) into (9) one finds the following estimate:
\begin{equation}
P\left( {\left\vert S\right\vert >\hbar ^{n}\triangle N}\right) <\left( {%
\alpha \vee _{0}/\triangle \vee }\right) ^{n}\ll 1.  \label{eq:12}
\end{equation}%
Applying the estimate (12) to (8) one can state that with the probability
close to unity the following relation holds:
\begin{equation}
\left( {\triangle \vee }\right) ^{n}\rho _{N}=\triangle N/N,  \label{eq:13}
\end{equation}%
signifying that the construction (7) plays the role of the probability
amplitude for the coordinate distribution of solitons' centers, with $\rho
_{N}$ in {(13)} being the corresponding probability density. Now let us
consider the measuring procedure for some observable $A$ corresponding, due
to E.~ Noether's theorem, to the symmetry group generator $\widehat{M}_{A}$.
For example, the momentum $\mathbf{P}$ is related with the generator of
space translation $\widehat{M}_{P}=-\imath \,\bigtriangledown $, the angular
momentum $\mathbf{L}$ is related with the generator of space rotation $%
\widehat{M}_{L}=\mathbf{J}$ and so on. As a result one can represent the
classical observable $A_{j}$ for the $j$--th trial in the form
\begin{equation*}
A_{j}=\int \,d^{3}x\,\pi _{j}\imath \widehat{M}_{A}\phi
_{j}=\sum_{k=1}^{n}\int \,d^{3}x\,\varphi _{j}^{\ast (k)}\widehat{M}%
_{A}^{(k)}\varphi _{j}^{(k)}.
\end{equation*}%
The corresponding mean value is
\begin{eqnarray}
\mathbb{E}(A) &\equiv &\frac{1}{N}\sum_{j=1}^{N}A_{j}=\frac{1}{N}%
\sum_{j=1}^{N}\sum_{k=1}^{n}\int \,d^{3}x\,\varphi _{j}^{\ast (k)}\widehat{M}%
_{A}^{(k)}\varphi _{j}^{(k)}  \notag \\
&=&\int \,d^{3n}x\,\Psi _{N}^{\ast }\widehat{A}\Psi _{N}+O(\frac{\vee _{0}}{%
\triangle \vee }),  \label{eq:14}
\end{eqnarray}%
where the Hermitian operator $\widehat{A}$ reads
\begin{equation}
\widehat{A}=\sum_{k=1}^{n}\hbar \widehat{M}_{A}^{(k)}.  \label{eq:15}
\end{equation}%
Thus, up to the terms of the order $\vee _{0}/\triangle \vee \ll 1$, we
obtain the standard quantum mechanical rule (14) for the calculation of mean
values. It is interesting to underline that the solitonian scheme in
question contains also the well-known spin---statistics correlation~[4].
Namely, if $\varphi _{j}^{(k)}$ is transformed under the rotation by
irreducible representation $D^{(J)}$ of $SO(3)$, with the weight $J$, then
the transposition of two identical extended particles is equivalent to the
relative $2\pi $--rotation of $\varphi _{j}^{(k)}$, that gives the
multiplication factor $(-1)^{2J}$ in $\Psi _{N}$. To show this property,
suppose that our particles are identical, i.e. their profiles $\varphi
_{j}^{(k)}$ may differ in phases only. Therefore, the transposition of the
particles with the centers at $\mathbf{r}_{1}$ and $\mathbf{r}_{2}$ means
the $\pi $--rotation of 2--particle configuration around the median axis of
the central vector line $\mathbf{r}_{1}-\mathbf{r}_{2}$. However, due to
extended character of the particles, to restore the initial configuration,
one should perform additional proper $\pi $--rotations of the particles. The
latter operation being equivalent to the relative $2\pi $--rotation of
particles, one concludes that it results in aforementioned multiplication of
$\Psi _{N}$ by $(-1)^{2J}$. Under the natural supposition that the weight $J$
is related with the spin of particles---solitons, one infers that the
many--particles wave function (7) should be symmetrical under the
transposition of the two identical particles if the spin is integer, but
antisymmetrical if the spin is half-integer (the Pauli principle). Thus, we
conclude that in the solitonian scheme the spin---statistics correlation
stems from the extended character of particles---solitons. However, the
particles in quantum mechanics being considered as point-like ones, it
appears inevitable to include the transpositional symmetry of the wave
function as the first principle (cf. Hartree---Fock receipt for Fermions).
Now it is worth-while to discuss the evidence of wave properties of
particles in solitonian scheme. To verify the fact that solitons can really
possess wave properties, the \textit{gedanken} diffraction experiment with
individual electrons---solitons was realized. Solitons with some velocity
were dropped into a rectilinear slit, cut in the impermeable screen, and the
transverse momentum was calculated which they gained while passing the slit,
with the width of the latter significantly exceeded the size of the soliton.
As a result, the picture of distribution of the centers of scattered
solitons was restored on the registration screen, by considering their
initial distribution to be uniform over the transverse coordinate. It was
clarified that though the center of each soliton fell into a definite place
of the registration screen (depending on the initial soliton profile and the
point of crossing the plane of the slit by the soliton's center), the
statistical picture in many ways was similar to the well-known diffraction
distribution in optics, i.e. the Fresnel's picture at short distances from
the slit and the Fraunhofer's one at large distances~[9].

\section{Random Hilbert space}

As a result we obtain the stochastic realization (7) of the wave function $%
\Psi _{N}$ which can be considered as an element of the random Hilbert space
$\mathcal{H}_{\mbox{rand}}$ with the inner product
\begin{equation}
(\psi _{1},\psi _{2})=\mathbb{M}(\psi _{1}^{\ast }\psi _{2}),  \label{eq:16}
\end{equation}%
with $\mathbb{M}$ standing for the expectation value. As a rude
simplification one can admit that the averaging in (16) is taken over random
characteristics of particles---solitons, such as their positions,
velocities, phases, and so on. It is important to underline once more that
the correspondence with the standard quantum mechanics is retained only in
the point--particle limit ($\triangle \vee \gg \vee _{0}$) for $N\rightarrow
\infty $. To show this~[2] one can apply the central limit theorem stating
that for $N\rightarrow \infty $\ the wave function $\Psi _{N}(t,\mathbf{x})$
behaves as the Gaussian random field with the variance
\begin{equation}
{\sigma }^{2}=\rho (t,\mathbf{x}),\qquad \mathbf{x}\in \mathbb{R}^{3n},
\label{eq:17}
\end{equation}%
where $\rho (t,\mathbf{x})$ stands for the probability density (partition
function) of solitons' centers in $\mathbb{R}^{3n}$. Random Hilbert spaces
being widely exploited in mathematical statistics, for quantum applications
they were first used by N.~Wiener in~[10]. To illustrate the line of
Wiener's argument, we recall the general scheme of introducing various
representations in quantum mechanics. Let $|\psi \rangle $~be a state vector
in the Hilbert space $\mathcal{H}$ and $\widehat{A}$~be a self-conjugate
operator with the spectrum $\sigma (\widehat{A})$. Then the $a$%
--representation is given by the wave function
\begin{equation*}
\psi (a)=\langle a|\psi \rangle ,
\end{equation*}%
where
\begin{equation*}
\widehat{A}|a\rangle =a|a\rangle ,\qquad a\in \sigma (\widehat{A}).
\end{equation*}%
In particular, the famous Schrodinger coordinate $q$--representation is
given by the wave function
\begin{equation}
\psi (q)=\langle q|\psi \rangle =\sum_{n}\,\langle q|n\rangle \langle n|\psi
\rangle ,  \label{eq:18}
\end{equation}%
with $|n\rangle $ being some complete set of state vectors in $\mathcal{H}$.
Wiener considered the real Brownian process $x(s,\alpha )$ in the interval $%
[0,1]\ni s$, where $\alpha \in \lbrack 0,1]$ is the generalized index of the
Brownian trajectory and the correlation reads
\begin{equation}
\int_{0}^{1}\,d\alpha \,x(s,\alpha )x(s^{\prime },\alpha )=\min
\,(s,s^{\prime }).  \label{eq:19}
\end{equation}%
To obtain the quantum mechanical description, Wiener defined the complex
Brownian process
\begin{equation}
z(s|\alpha ,\beta )=\frac{1}{\sqrt{2}}\left[ x(s,\alpha )+\imath \,y(s,\beta
)\right] ;\quad \alpha ,\,\beta \in \lbrack 0,1],  \label{eq:20}
\end{equation}%
and using the natural mapping $\mathbb{R}^{3}\rightarrow \lbrack 0,1]$, for
the particle in $\mathbb{R}^{3}$, constructed the stochastic representation
of the wave function along similar lines as in (18):
\begin{equation}
\langle \alpha ,\beta |\psi \rangle =\int_{s\in \lbrack 0,1]}\,dz(s|\alpha
,\beta )\psi (s),  \label{eq:21}
\end{equation}%
with the obvious unitarity property
\begin{equation*}
\int_{0}^{1}\,ds\,|\psi (s)|^{2}=\iint\limits_{[0,1]^{2}}\,d\alpha \,d\beta
|\langle \alpha ,\beta |\psi \rangle |^{2}
\end{equation*}%
stemming from (19).

\section{Entangled solitons and EPR correlations}

In the sequel we shall consider the special case of two--particle
configurations ($n=2$), corresponding to the singlet state of two spin--$1/2$
particles. In quantum mechanics these states are described by the spin wave
function of the form
\begin{equation}
\psi _{12}=\frac{1}{\sqrt{2}}\left( |1\,\uparrow \rangle \otimes
|2\,\downarrow \rangle -|1\,\downarrow \rangle \otimes |2\,\uparrow \rangle
\right)   \label{eq:22}
\end{equation}%
and are known as \textbf{entangled states}. The arrows in (22) signify the
projections of spin $\pm 1/2$ along some fixed direction. In the case of the
electrons in the famous Stern---Gerlach experiment this direction is
determined by that of an external magnetic field. If one chooses two
different Stern---Gerlach devices, with the directions $\mathbf{a}$ and $%
\mathbf{b}$ of the magnetic fields, denoted by the unit vectors $\mathbf{a}$
and $\mathbf{b}$ respectively, one can measure the correlation of spins of
the two electrons by projecting the spin of the first electron on $\mathbf{a}
$ and the second one on $\mathbf{b}$. Quantum mechanics gives for the spin
correlation function the well-known expression
\begin{equation}
P(\mathbf{a},\,\mathbf{b})=\psi _{12}^{+}(\mathbf{\sigma a})\otimes (\mathbf{%
\sigma b})\psi _{12},  \label{eq:23}
\end{equation}%
where $\mathbf{\sigma }$ stands for the vector of Pauli matrices $\sigma _{i}
$, $i=1,2,3$. Putting (22) into (23), one easily gets
\begin{equation}
P(\mathbf{a},\,\mathbf{b})=-(\mathbf{ab}).  \label{eq:24}
\end{equation}%
The formula (24) characterizes the spin correlation in the
Einstein---Podolsky---Rosen entangled singlet states and is known as the
EPR--correlation. As was shown by J.~Bell~[11], the correlation (24) can be
used as an efficient criterium for distinguishing the models with the local
(point-like) hidden variables from those with the nonlocal ones. Namely, for
the local-hidden-variables theories the EPR--correlation (24) is broken. It
would be interesting to check the solitonian model shortly described in the
beforehand points by applying to it the EPR--correlation criterium. To this
end let us first describe the spin--$1/2$ particles as solitons in the
nonlinear spinor model of Heisenberg---Ivanenko type considered in the
works~[12, 13]. The soliton in question is described by the relativistic
4--spinor field $\varphi $ of stationary type
\begin{equation}
\varphi =\left[
\begin{array}{c}
u \\
v%
\end{array}%
\right] \,e^{-\imath \omega t},  \label{eq:25}
\end{equation}%
satisfying the equation
\begin{equation}
\left( \imath \gamma ^{k}\partial _{k}-\ell _{0}^{-1}+\lambda (\overline{%
\varphi }\varphi )\right) \varphi =0,  \label{eq:26}
\end{equation}%
where $u$ and $v$ denote 2--spinors, $k$ runs Minkowsky space indices 0, 1,
2, 3; $\ell _{0}$ stands for some characteristic length (the size of the
particle---soliton), $\lambda $ is the self-coupling constant, $\overline{%
\varphi }\equiv \varphi ^{+}\gamma ^{0}$, $\gamma ^{k}$ are the Dirac
matrices. The stationary solution to the equation (26) can be obtained by
separating variables in spherical coordinates $r$, $\vartheta $, $\alpha $
via the substitution
\begin{equation}
u=\frac{1}{\sqrt{4\pi }}f(r)\left[
\begin{array}{c}
1 \\
0%
\end{array}%
\right] ,\quad v=\frac{\imath }{\sqrt{4\pi }}g(r)\sigma _{r}\left[
\begin{array}{c}
1 \\
0%
\end{array}%
\right] ,  \label{eq:27}
\end{equation}%
where $\sigma _{r}=(\mathbf{\sigma r})/r$. Inserting (27) into (26) one
finds
\begin{gather*}
\frac{\omega }{c}u+\imath (\mathbf{\sigma \bigtriangledown })v-\ell
_{0}^{-1}u+\frac{\lambda }{4\pi }\left( f^{2}-g^{2}\right) u=0, \\
\frac{\omega }{c}v+\imath (\mathbf{\sigma \bigtriangledown })u-\ell
_{0}^{-1}v+\frac{\lambda }{4\pi }\left( f^{2}-g^{2}\right) v=0.
\end{gather*}%
In view of (27) one gets
\begin{gather*}
\imath (\mathbf{\sigma \bigtriangledown })v=-\frac{1}{\sqrt{4\pi }}\left(
g^{\prime }+\frac{2}{r}g\right) \left[
\begin{array}{c}
1 \\
0%
\end{array}%
\right] , \\
\imath (\mathbf{\sigma \bigtriangledown })u=-\frac{\imath }{\sqrt{4\pi }}%
f^{\prime }\sigma _{r}\left[
\begin{array}{c}
1 \\
0%
\end{array}%
\right] .
\end{gather*}%
Finally, one derives the following ordinary differential equations for the
radial functions $f(r)$ and $g(r)$:
\begin{gather*}
\left( g^{\prime }+\frac{2}{r}g\right) =\left( \frac{\omega }{c}-\ell
_{0}^{-1}\right) f+\frac{\lambda }{4\pi }\left( f^{2}-g^{2}\right) f, \\
-f^{\prime }=\left( \frac{\omega }{c}+\ell _{0}^{-1}\right) g+\frac{\lambda
}{4\pi }\left( f^{2}-g^{2}\right) g.
\end{gather*}%
As was shown in the papers~[12, 13], these equations admit regular solutions
if the frequency parameter $\omega $ belongs to the interval
\begin{equation}
0<\omega <c/{\ell _{0}}.  \label{eq:28}
\end{equation}%
The behavior of the functions $f(r)$ and $g(r)$ at $r\rightarrow 0$ is as
follows:
\begin{equation*}
g(r)=C_{1}r,\quad f=C_{2},\quad f^{\prime }\rightarrow 0,
\end{equation*}%
where $C_{1}$, $C_{2}$ denote some integration constants. The behavior of
solutions far from the center of the soliton, i.e. at $r\rightarrow \infty $%
, is given by the relations:
\begin{equation*}
f=\frac{A}{r}e^{-\nu r},\quad g=-\frac{f^{\prime }}{B},
\end{equation*}%
where
\begin{equation*}
\nu =\left( \ell _{0}^{-2}-{\omega ^{2}}/{c^{2}}\right) ^{1/2},\quad B=\ell
_{0}^{-1}+{\omega }/c.
\end{equation*}%
If one chooses the free parameters $\ell _{0}$ and $\lambda $ of the model
to satisfy the normalization condition (similar to (6))
\begin{equation}
\int d^{3}x\,\varphi ^{+}\varphi =\int\limits_{0}^{\infty }dr\,r^{2}\left(
f^{2}+g^{2}\right) =\hbar   \label{eq:29}
\end{equation}%
then the spin of the soliton reads
\begin{equation}
\mathbf{S}=\int d^{3}x\,\varphi ^{+}\mathbf{J}\varphi =\frac{\hbar }{2}%
\mathbf{e}_{z},  \label{eq:30}
\end{equation}%
where $\mathbf{e}_{z}$ denotes the unit vector along the $Z$--direction, $%
\mathbf{J}$ stands for the angular momentum operator
\begin{equation}
\mathbf{J}=-\imath \lbrack \mathbf{r}\mathbf{\bigtriangledown }]+\frac{1}{2}%
\,\mathbf{\sigma }\otimes \sigma _{0},  \label{eq:31}
\end{equation}%
and $\sigma _{0}$ is the unit $2\times 2$--matrix. Now let us construct the
two--particles singlet configuration on the base of the soliton solution
(25). First of all, in analogy with (22) one constructs the entangled
solitons configuration endowed with the zero spin:
\begin{equation}
\varphi _{12}=\frac{1}{\sqrt{2}}\left[ \varphi _{1}^{\uparrow }\otimes
\varphi _{2}^{\downarrow }-\varphi _{1}^{\downarrow }\otimes \varphi
_{2}^{\uparrow }\right] ,  \label{eq:32}
\end{equation}%
where $\varphi _{1}^{\uparrow }$ corresponds to (27) with $\mathbf{r}={%
\mathbf{r}}_{1}$, and $\varphi _{2}^{\downarrow }$ emerges from the above
solution by the substitution
\begin{equation*}
{\mathbf{r}}_{1}\rightarrow {\mathbf{r}}_{2},\quad \left[
\begin{array}{c}
1 \\
0%
\end{array}%
\right] \rightarrow \left[
\begin{array}{c}
0 \\
1%
\end{array}%
\right] \,
\end{equation*}%
that corresponds to the opposite projection of spin on the $Z$--axis. In
virtue of the orthogonality relation for the states with the opposite spin
projections one easily derives the following normalization condition for the
entangled solitons configuration (32):
\begin{equation}
\int \,d^{3}x_{1}\,\int \,d^{3}x_{2}\,\varphi _{12}^{+}\varphi _{12}={\hbar }%
^{2}.  \label{eq:33}
\end{equation}%
Now it is not difficult to find the expression for the stochastic wave
function (7) for the singlet two--solitons state:
\begin{equation}
\Psi _{N}\left( t,{\mathbf{r}}_{1},{\mathbf{r}}_{2}\right) ={\left( {\hbar }%
^{2}N\right) }^{-1/2}\sum\limits_{j=1}^{N}\,\varphi _{12}^{(j)},
\label{eq:34}
\end{equation}%
where $\varphi _{12}^{(j)}$ corresponds to the entangled soliton
configuration in the $j$--th trial. Our next step is the calculation of the
spin correlation (23) for the singlet two--soliton state. In the light of
the fact that the operator $\mathbf{\sigma }$ in (23) corresponds to the
twice angular momentum operator (31) one should calculate the following
expression:
\begin{equation}
P^{\prime }(\mathbf{a},\mathbf{b})=\mathbb{M}\,\int \,d^{3}x_{1}\,\int
\,d^{3}x_{2}\,\Psi _{N}^{+}2\left( {\mathbf{J}}_{1}\mathbf{a}\right) \otimes
2\left( {\mathbf{J}}_{2}\mathbf{b}\right) \Psi _{N},  \label{eq:35}
\end{equation}%
where $\mathbb{M}$ stands for the averaging over the random phases of the
solitons. Inserting (34) and (31) into (35), using the independence of
trials $j\not=j^{\prime }$, and taking into account the relations:
\begin{align*}
J_{+}\varphi ^{\uparrow }& =0, & J_{3}\varphi ^{\uparrow }& =\frac{1}{2}%
\varphi ^{\uparrow }, & J_{-}\varphi ^{\uparrow }& =\varphi ^{\downarrow },
\\
J_{-}\varphi ^{\downarrow }& =0, & J_{3}\varphi ^{\downarrow }& =-\frac{1}{2}%
\varphi ^{\downarrow }, & J_{+}\varphi ^{\downarrow }& =\varphi ^{\uparrow },
\end{align*}%
where $J_{\pm }=J_{1}\pm \imath J_{2}$, one easily finds that
\begin{equation}
P^{\prime }(\mathbf{a},\mathbf{b})=-{\hbar }^{-2}\left( \mathbf{a}\mathbf{b}%
\right) \left( \int\limits_{0}^{\infty }\,dr\,r^{2}\left( f^{2}+g^{2}\right)
\right) ^{2}=-\left( \mathbf{a}\mathbf{b}\right) .  \label{eq:36}
\end{equation}%
Comparing the correlations (36) and (24) one remarks their coincidence, that
is the solitonian model satisfies the EPR--correlation criterium.

\section{Conclusion. Simulation of qubits by probabilistic bits}

Now we intend to explain how stochastic qubits introduced previously could
be simulateded by standard probabilistic bits~[14]. To this end one should
define the random phase $\Phi _{j}$ for the $j$--th trial in our system of $n
$ solitons---particles. Let ${\varphi }^{(k)}(\mathbf{r})$ denote the
standard (etalon) profile for the $k$--th soliton. The most probable
position $\mathbf{d}_{j}^{(k)}(t)$ of the $k$--th soliton's center for $j$%
--th trial can be found from the following variational problem:
\begin{equation*}
\left\vert \int \,d^{3}x\,{\varphi }_{j}^{\ast (k)}(t,\mathbf{r}){\varphi }%
^{(k)}\left( \mathbf{r}-\mathbf{d}_{j}^{(k)}\right) \right\vert \quad
\rightarrow \quad \max ,
\end{equation*}%
thus giving the random phase structure:
\begin{equation}
\Phi _{j}=\sum_{k=1}^{n}\,\text{arg}\,\int \,d^{3}x\,{\varphi }_{j}^{\ast
(k)}(t,\mathbf{r}){\varphi }^{(k)}\left( \mathbf{r}-\mathbf{d}%
_{j}^{(k)}\right) .  \label{eq:37}
\end{equation}%
The random phase (37) can be used for simulating quantum computing via
generating the following $K$ random dichotomic functions:
\begin{equation}
f_{s}(\theta _{s})=\mbox{sign}\,\left[ \cos (\Phi _{j}+\theta _{s})\right]
,\qquad s=\overline{1,K},  \label{eq:38}
\end{equation}%
with $\theta _{s}$ being arbitrary fixed phases. Now recall that the qubit
is identified with the state vector
\begin{equation*}
|\psi \rangle =\alpha |0\rangle +\beta |1\rangle ,
\end{equation*}%
corresponding to the superposition of two orthogonal states $|0\rangle $ and
$|1\rangle $, as for instance, two polarizations of the photon, or two
possible spin--$1/2$ states. It is worth-while to compare the standard
EPR--correlation (24) with the random phases one for the case of $n=2$
particles:
\begin{equation}
\mathbb{E}\,(f_{1}\,f_{2})=1-\frac{2}{\pi }|\triangle \theta |,
\label{eq:39}
\end{equation}%
where $\triangle \theta =\theta _{1}-\theta _{2}$. The similarity of these
two functions (39) and (24) of the angular variable seems to be a good
motivation for the $K$ qubits simulation by the dichotomic random functions
(38) popularized in the paper~[15]. This very simple model of stochastic
qubits simulation can be employed for simulating Bi-photons, EPR states and
other entanglement states. We hope that this model will be useful for the
Shor's and Grover's Quantum Algorithms realization.

\subsection{Initialization of probabilistic bits}

Let us consider $N$ stochastic qubits constructed according to the above
algorithm. Let us call qubit No. 1 the control or signal one. If the signal
qubit is green then the value assigned will be $\left| 0\right\rangle $, if
it is red the output will be $\left| 1\right\rangle $. Hence, at each output
we shall get $N$ initialized integral qubits.

\subsection{Hadamard transform for probabilistic bits}

Let us consider the initial bit
\begin{equation*}
\left| q\right\rangle =\frac{1}{\sqrt{2}}(\left| 0\right\rangle +\left|
1\right\rangle ),
\end{equation*}
with equal probabilities $P = 1/2$ being assigned to the states $\left|
0\right\rangle $ and $\left| 1\right\rangle $. Consider now the Hadamard
transform $H$ sending the states to the basis rotated over $\pi/4$:
\begin{equation*}
H\left| 0\right\rangle \rightarrow \frac{1}{\sqrt{2}}(\left| 0\right\rangle
+\left| 1\right\rangle ), \quad H\left| 1\right\rangle \rightarrow \frac{1}{%
\sqrt{2}}(\left| 0\right\rangle -\left| 1\right\rangle ).
\end{equation*}
Hence, having applied the Hadamard transform to the bit $\left|q\right%
\rangle $ one gets the final bit
\begin{equation*}
H\left| q\right\rangle =\left| 0\right\rangle .
\end{equation*}

\subsection{Logical $CNOT$ for probabilistic bits}

The result of application of the logical component $CNOT$ depends on the
target bit, two possible cases being considered:\newline
\noindent 1. If the control bit has the value $\left\vert 1\right\rangle $
then the target bit is sent to the opposite value.\newline
\noindent 2. If the control bit has the value $\left\vert 0\right\rangle $
then the value of the target bit is not changed.\newline
All these elementary operations can be realized via classical computer
through simulating the phase structure of realistic solitons by the
generator of random numbers connected to the model solitons' generator, e.
g. Kerr dielectric with the optical excitations or magnetic with the
excitations of localized spin inversion domains.

\end{document}